\documentclass[runningheads]{llncs}
\usepackage{graphicx}
\usepackage{verbatim}
\usepackage{float}
\begin{document}
\title{Multi Scale Supervised 3D U-Net for Kidney and Tumor Segmentation\thanks{Supported by Fudan University.}}
%
%
\author{Wenshuai Zhao \and
Zengfeng Zeng}
\authorrunning{W. Zhao and Z. Zeng}
%
\institute{Department of Electronic Engineering, Fudan University, Shanghai, China
\\
}
\maketitle              
\begin{abstract}
U-Net has achieved huge success in various medical image segmentation challenges. Kinds of new architectures with bells and whistles might succeed in certain dataset when employed with optimal hyperparameter, but their generalization always can’t be guaranteed. Here, we focused on the basic U-Net architecture and proposed a multi scale supervised 3D U-Net for the segmentation task in KiTS19 challenge. To enhance the performance, our work can be summarized as three folds: first, we used multi scale supervision in the decoder pathway, which could encourage the network to predict right results from the deep layers; second, with the aim to alleviate the bad effect from the sample imbalance of kidney and tumor, we adopted exponential logarithmic loss; third, a connected-component based post processing method was designed to remove the obviously wrong voxels. In the published KiTS19 training dataset (totally 210 patients), we divided 42 patients to be test dataset and finally obtained DICE scores of 0.969 and 0.805 for the kidney and tumor respectively. In the challenge, we finally achieved the 7th place among 106 teams with the Composite Dice of 0.8961, namely 0.9741 for kidney and 0.8181 for tumor. 

\end{abstract}
\section{Introduction}
Automatic semantic segmentation provides a promising tool for exploring the relationship between tumor morphology and its corresponding surgical outcome, as well as developing advanced surgical planning techniques \cite{ref_1,ref_2,ref_3}, but it is still challenging to achieve good performance due to the morphological heterogeneity.

The KiTS19 challenge \cite{ref_4} aims to accelerate the development of reliable kidney and kidney tumor semantic segmentation methodologies. It has CT scans of 300 unique kidney cancer patients and releases 210 of these for model training and validation, the other 90 patient scans will be reserved as test dataset. The final performance of the submitted model is measured based on the mean DICE coefficient of kidney and tumor segmentation.

Deep convolution neural networks (CNNs) have been recognized as the state-of-the-art method for various image classification and segmentation tasks. U-Net \cite{ref_5} with the encoder-decoder architecture is a stably successful network for medical image segmentation. As volumetric data are more abundant than 2D images in biomedical data analysis such, to fully utilize the spatial information of 3D images such as CT and MRI, 3D convolution is proposed and thought to be much more effective. Fabian\cite{ref_6} did minor modification based on 3D U-Net \cite{ref_7} and got champions or top ranks in many medical image segmentation challenges, which proves that an optimized U-Net possesses enough potential to perform better than many other new architectures.

Inspired by \cite{ref_6} we also discarded some common tricks on architecture, such as residual block \cite{ref_8}, dense block \cite{ref_9}, attention mechanism \cite{ref_10}, feature pyramid network \cite{ref_11} and feature recalibration \cite{ref_12}. From the point of our view, medical images are far less diverse than nature images, so they don’t need too deep convolution layers or too much connections. The basic U-Net with only 5 layers is enough to represent or learn the features to be used to classify.

Following such suggestion, we focused our work on training 3D U-Net better and utilizing the limited training dataset more effectively. As the final full resolution prediction is upsampled from deeper low-resolution layers, it is very important to guarantee the accuracy of the prediction in deep layers. So, we designed multi scale supervised 3D U-Net to encourage the network to predict right not only in the last layer, but also in every resolution level, which improved the performance in the final layer consequently. To mitigate the negative effect brought by imbalanced class data, we used the enhanced focal loss \cite{ref_13}, exponential logarithmic loss \cite{ref_14}. At last, our post processing would remove the scattered kidney or tumor not attached with kidney.

\section{Method}
In this part, we will present our method details, not only the network architecture but also including the preprocessing, data augmentation, training procedure, inference and post processing, because these are also very important to achieve the performance that 3D U-Net should have.

\subsection{Preprocessing and Data Augmentation}
To remove the abnormal intensity values, which might be from some metal things, we clipped the intensity values of CT images into their 0.5 and 99.5th percentile. Then by convention, we normalized the data with global foreground mean and standard deviation due to the typical weight initialization method. It should be emphasized that the anisotropy of 3D data would destroy the advantage of 3D convolution since it cannot learn a unified representation for different voxel space data with the same receptive field. Thus, we resampled the data into a same voxel space if they are not.

It is always tedious to manually annotate medical images, so the labeled datasets volume is usually limited. We employed strong data augmentation to avoid the model being overfitted, including random rotations, random scaling, random elastic deformations, gamma correction augmentation and mirroring.

\subsection{Network Architecture}
U-Net [5] is a classical encoder-decoder segmentation network and has drawn a lot of attention in recent years. The encoder pathway is similar with the typical classification network to extract more high-level semantic feature layer by layer. Then the decoder pathway recovers the localization for every voxel and utilizes the feature information to classify it. To employ the position information embedded in encoder, direct connection is constructed between the layers in the same stage.

We designed our network based on 3D U-Net [7]. The framework is shown as Fig. 1. This multi scale supervised network makes prediction from different layers of decoder pathway, unlike classical 3D U-Net only predicts from the last layer. These segment outputs would be compared with corresponding resolution labels and then used to calculate the final loss function. Such supervision encourages the network to predict correctly from the low-resolution feature maps which will be up-sampled to be full resolution feature maps.

\begin{figure}
	\includegraphics[width=\textwidth]{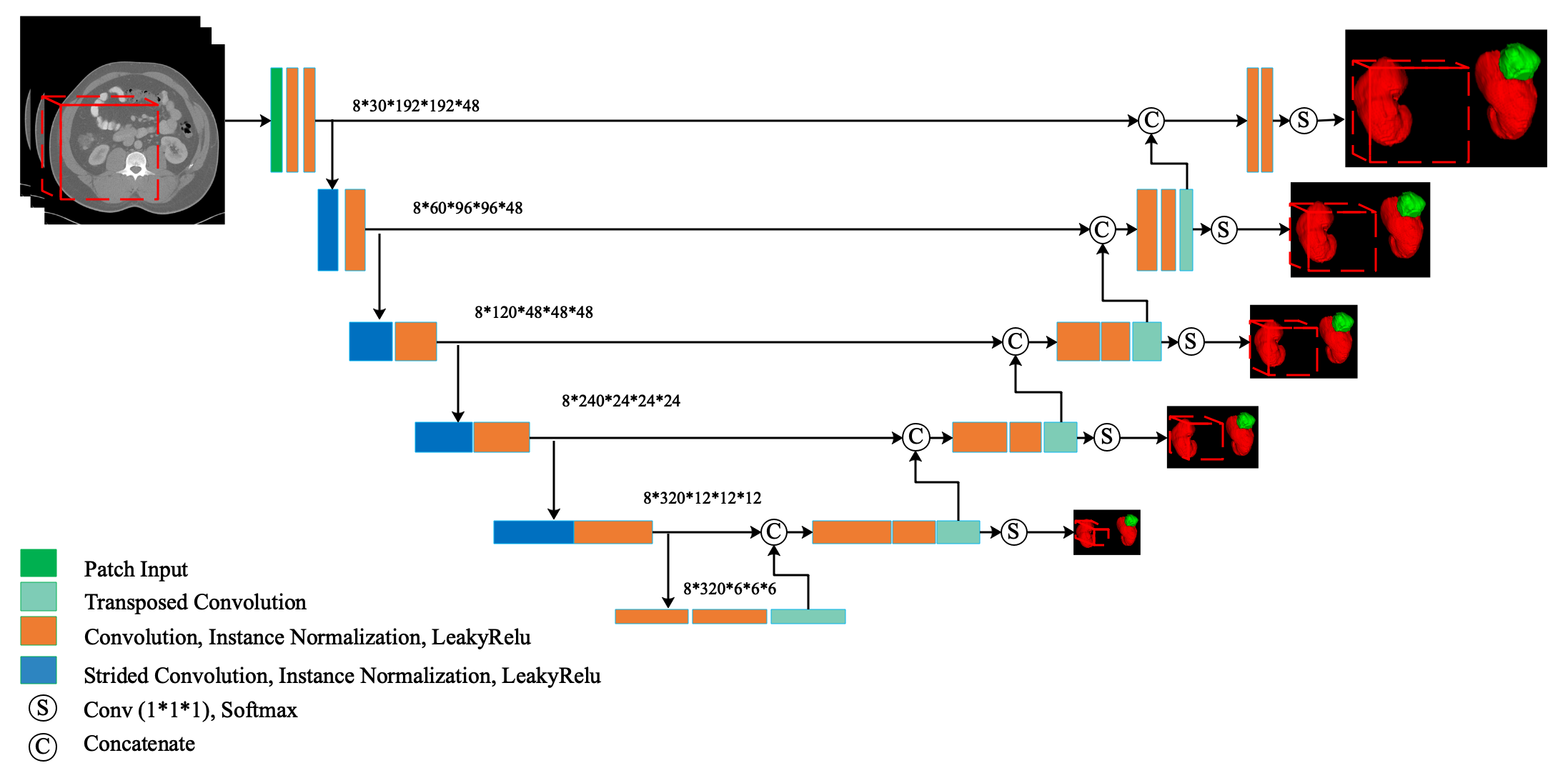}
	\caption{The scheme of our multi scale supervised 3D U-Net (best viewed in color). The actual architecture is 3D, but for simplicity we use 2D here. From our experience, we adopt strided convolution instead of pooling operation and replace trilinear interpolation by transposed convolution to up sample. To reduce the model volume, we set the basic feature number as 30.} \label{fig1}
\end{figure}
\subsection{Training Procedure}
Limited by the GPU memory, we chose the patch size as $192\times{192}\times{48}$ and set the batch size as 8 using Data Parallel in 2 GPUs (Tesla, 32GB). The patch was random sampled and we refer to an epoch as 250 iterations. We used Adam as our optimizer. The learning rate was initialized to be $3\times10^{-4}$, and would be dropped by the factor of 0.2 if the training loss was no more improved in 30 epochs. 

For kidney and tumor segmentation in CT images, the sample of background is far more than kidney and tumor voxels. Also, the tumor is more difficult to classify due to its morphological heterogeneity. In order to mitigate this imbalance, we used exponential logarithmic loss \cite{ref_14}. This loss emphasizes the effect of diffcult samples and gives them more weight by making the loss nonlinear.  At the same time, we attributed different weights for background, kidney and tumor manually.

We combined the Soft Dice and Cross Entropy to train our model. The final fromat of our loss can be summarized as follows:

\begin{equation}
Loss=DICE+CE
\end{equation}
\begin{equation}
DICE={(-{log^{Dice_{kidney}}})}^{0.3}\times{0.4}+{(-{log^{Dice_{tumor}}})}^{0.3}\times{0.6}
\end{equation}
\begin{equation}
CE={0.28\times{CE_{bg}}}+{0.28\times{CE_{kidney}}}+{0.44\times{CE_{tumor}}}
\end{equation}

\subsection{Inference and Post Processing}
When cases are predicted, we used a sliding window approach and there are overlaps between predictions. To improve the accuracy, the predictions from original data and mirrored data were combined.
\begin{figure}
	\includegraphics[width=\textwidth]{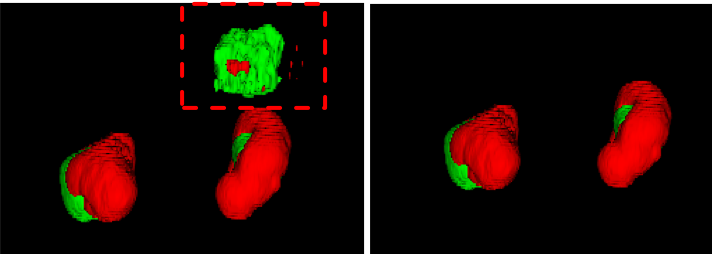}
	\caption{The effect of our post processing method. The left image is before post processed, and the volume in dotted box are some voxels obviously wrong. The right image is after post processed, and the extra voxels have been removed.} \label{fig2}
\end{figure}

\noindent Some common human knowledge could help to enhance the performance further. For example, there are at most two kidneys in one patient and the tumor should be attached with kidney. So, we designed a simple connected-component based post processing method to remove the obviously mistaken predictions. The effect is shown as Fig.2.

\section{Experiments and Results}
There are 210 patient CT scans published to train models. We divided 42 of these to be the test dataset and used other images to train our model. It consumed about 5 days running on 2 GPUs (Tesla 32GB). The loss during training is shown as Fig. 3. Stable reduction of our proposed loss could be obversed, and this process continued for about 700 epoches when the learning rate got the end of our patience. 

\begin{figure}
	\centering
	\includegraphics[width=0.9\textwidth]{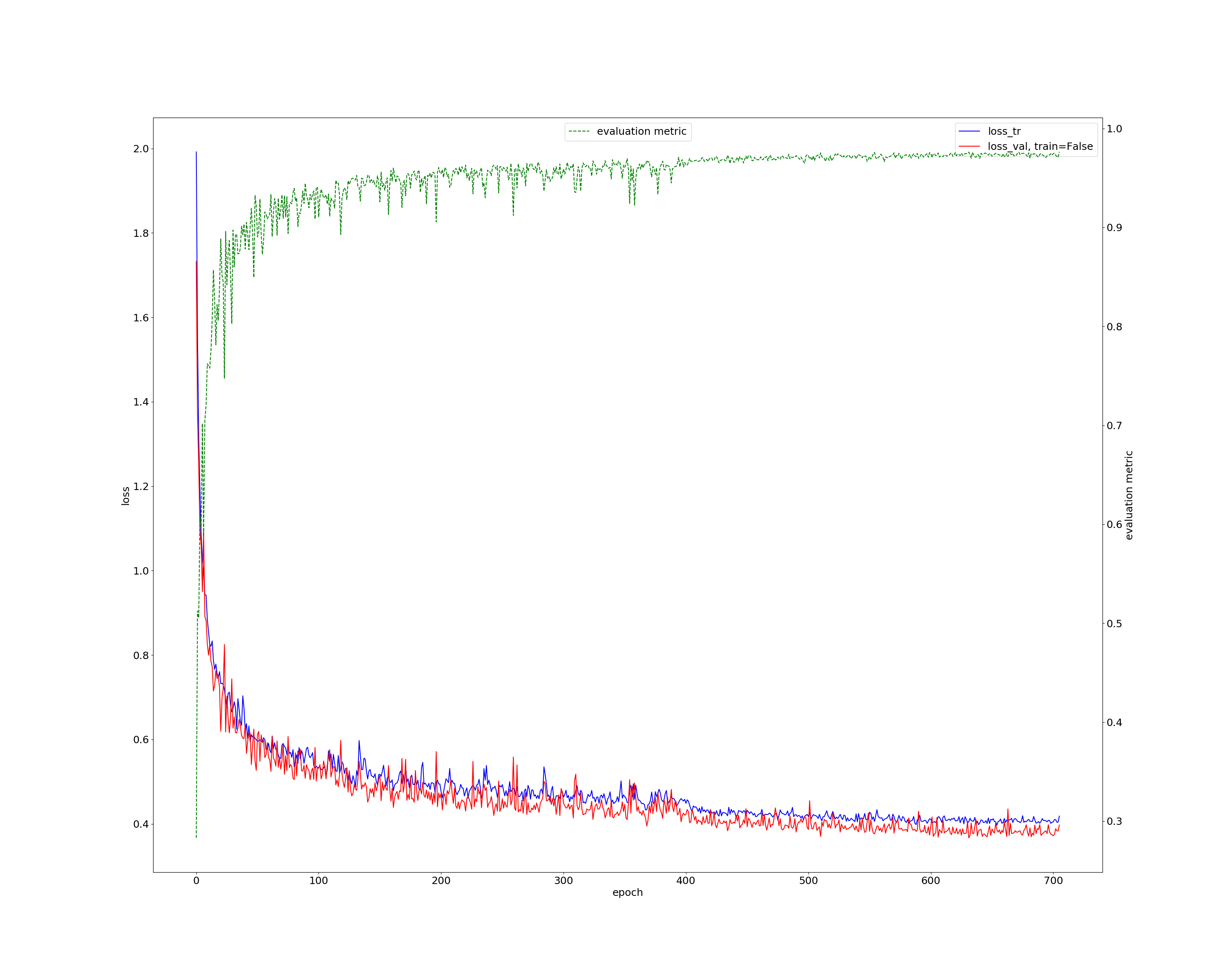}
	\caption{The loss changing during our training. Red and blue lines mean validation and training loss respectively. The green line is a sliding validation loss metric to choose the best check point.} \label{fig3}
\end{figure}

\noindent The samples of our segmentation outputs are shown in Fig.4. We observed both 2D slices from different views and 3D view to analyze the performance. Obviously, the kidney in CT images was segmented pretty well. The prediction was closely approximate the ground truth. However, some tumor was too small to get good Dice Coefficient, and also was difficult to find. Some cases with small tumor size would decrease the average Dice Coefficient drastically.

\begin{figure}[H]
	\centering
	\includegraphics[width=0.9\textwidth]{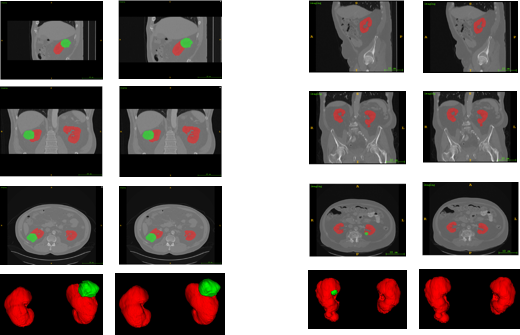}
	
	\caption{Two samples of our segmentation outputs. The rows from up to bottom are sagittal, coronal, transverse plane in 2D and the 3D view (best viewed in color). The columns from left to right are the ground truth and prediction of one common case and the worst case.} \label{fig4}
\end{figure}

\begin{figure}[H]
	\centering
	\includegraphics[width=0.85\textwidth]{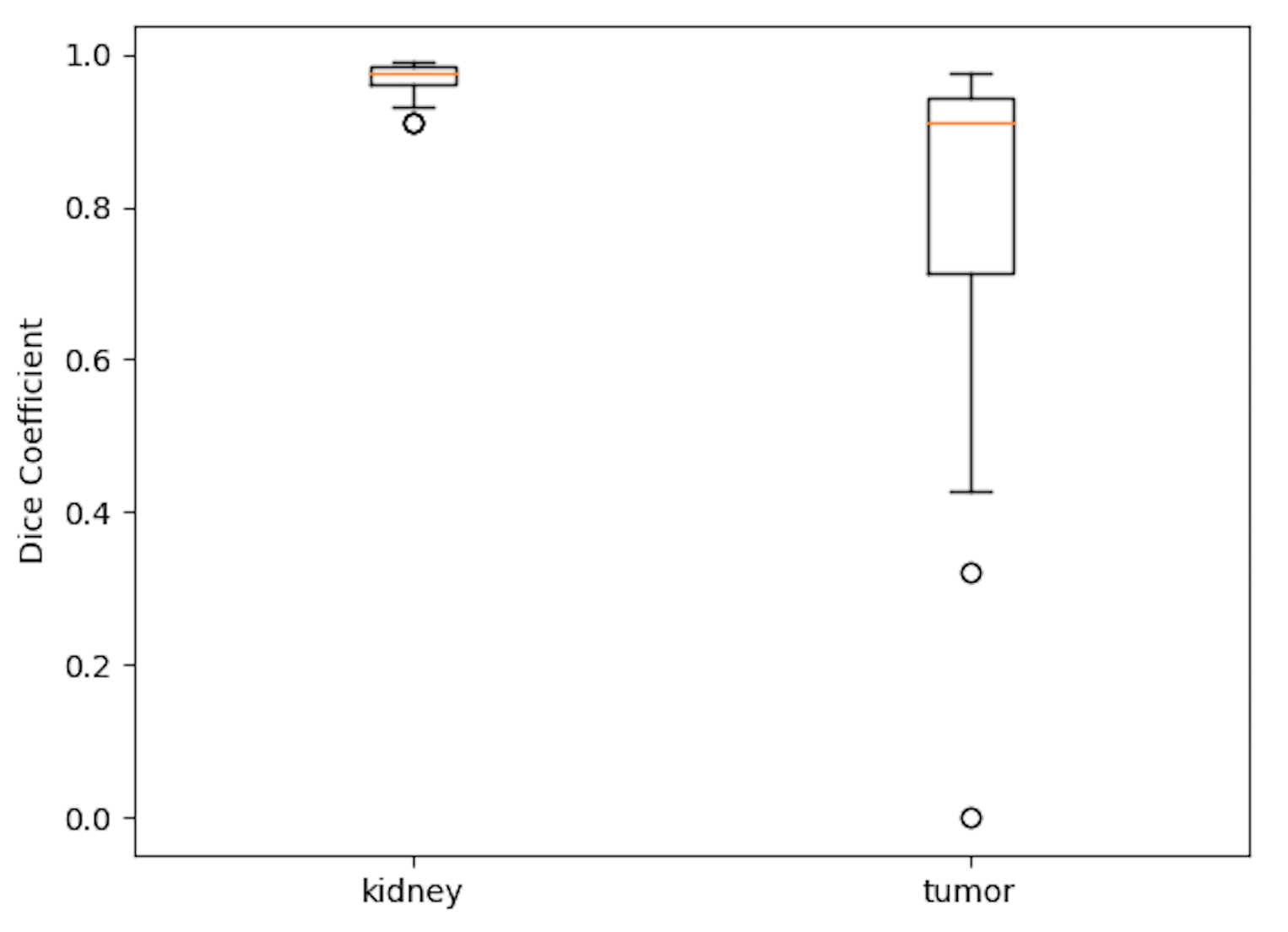}
	\caption{The boxplot of the segmentation outputs for the 42 patients in divided test dataset.} \label{fig5}
\end{figure}
\noindent The boxplot of our test dataset is shown in Fig. 5. The average Dice Coefficient of test dataset are 0.969 and 0.805 for kidney and tumor respectively. Also, the varience of kidney is very small indicating that our algorithm is stable for kidney segmentation.

\section{Conclusion and Discussion}
In this paper we demonstrated our model and its performance in KiTS19 dataset. Our method was built based on classical 3D U-Net, and enhanced by multi scale supervision, exponential logarithmic loss, and connected-component based post processing. Tested in the published data, our method achieved average Dice Coefficient of 0.969 and 0.805 for kidney and tumor respectively. 

And in the challenge, we got Composite Dice of 0.8961 with 0.9741 for kidney and 0.8181 for tumor, which ranking in the $7^{th}$ place among all the 106 teams. The Dice Coefficient of kdney segmentation has slight margin to the highest 0.9743, but the one of tumor segmentation is outperformed by the top teams.

Since we followed the spirit of Fabian \cite{ref_6}, we did not use too much architecture tricks here, but focused on the training procedure. Tumor with serious morphological heterogeneity is always hard to segment well. In the future, we plan to try two-stage method for tumor segmentation, that is we should propose the region of interest first and then use deformable convolution instead of conventional convolution to suit the tumor feature.

\end{document}